\documentclass{article}

\usepackage{longtable}
\usepackage{booktabs}
\usepackage[numbers,sort&compress]{natbib}
\PassOptionsToPackage{options}{natbib}
\usepackage[preprint]{neurips_2026}
\usepackage{booktabs}
\usepackage[table]{xcolor}
\usepackage[most]{tcolorbox}
\usepackage{booktabs}
\usepackage{xcolor}
\usepackage{colortbl}
\usepackage{amsmath}
\usepackage{listings}
\usepackage{xcolor}
\usepackage{graphicx}
\usepackage{subcaption}
\usepackage{xcolor}
\usepackage{tikz}
\usepackage{enumitem}

\usepackage[most]{tcolorbox}
\tcbuselibrary{breakable,listings,skins}

\newcommand{\stepbadge}[1]{%
\tikz[baseline=(char.base)]{
\node[
    fill=brown!120,
    text=white,
    rounded corners=1pt,
    inner xsep=3.5pt,
    inner ysep=1pt,
    font=\bfseries\normalsize
] (char) {#1};
}%
}
\usepackage[utf8]{inputenc} 
\usepackage[T1]{fontenc}    
\usepackage{hyperref}       
\usepackage{url}            
\usepackage{booktabs}       
\usepackage{amsfonts}       
\usepackage{nicefrac}       
\usepackage{microtype}      
\usepackage{xcolor}         

\title{Neurosymbolic Auditing of Natural-Language Software Requirements}

\author{
  Bethel Hall \\
  Stevens Institute of Technology \\
  \texttt{bhall2@stevens.edu}
  \And
  William Eiers \\
  Stevens Institute of Technology \\
  \texttt{weiers@stevens.edu}
}

\begin{document}
\maketitle
\begin{abstract}
Natural-language software requirements are often ambiguous, inconsistent, and underspecified; in safety-critical domains, these defects propagate into formal models that verify the wrong specification and into implementations that ship unsafe behavior. We show that large language models, equipped with an SMT solver, can audit such requirements: translating them into formal logic, detecting ambiguity through stochastic variation in the generated formalization, and exposing inconsistency, vacuousness, and safety violations through solver queries on the resulting specification. We present \textsc{VeriMed}, a neurosymbolic pipeline that operationalizes this idea for medical-device software requirements, and report two findings. First, stochastic variation across independent formalizations is a signal of ambiguity: requirements that admit multiple plausible interpretations produce SMT-inequivalent formalizations, and bidirectional SMT equivalence checking turns this disagreement into a solver-checkable test. Second, the usefulness of symbolic feedback depends on its granularity: in counterexample-guided repair on a hemodialysis question-answering benchmark, concrete SMT counterexamples raise verified accuracy from 55.4\% to 98.5\%.
Over an extensive experimental evaluation on open-source hemodialysis safety requirements, we show that the LLM-based approach in \textsc{VeriMed} successfully reduces ambiguity-sensitive requirements and enables rigorous auditing of software requirements through SMT-based queries.

\end{abstract}

\section{Introduction}
Autoformalization, the translation of natural-language requirements into
machine-verifiable specifications, has advanced rapidly with large language
models~\citep{wu2022autoformalization}. Yet for safety-critical software
requirements, syntactic validity is not enough. A well-formed formal
specification can be globally inconsistent, contain requirements whose
trigger conditions are unreachable, or admit multiple plausible readings of
the same natural-language source. These failures are invisible by construction: when no ground-truth formalization exists, there is no oracle for correctness. This is inherent to LLM-generated formalizations, where the model resolves ambiguity by defaulting to the most statistically common reading in its training distribution or omits constraints implicit in the source.


This problem is especially acute in domains such as medical devices, 
where requirements specify alarm conditions, pump behavior, and safe parameter ranges. If the requirement set is
inconsistent, or if the
autoformalizer silently picks one reading among several, downstream verification
can succeed while checking the wrong specification entirely. To address this, we propose \textsc{VeriMed}, a neurosymbolic framework that makes these failures detectable through two
complementary checks. First, we audit the generated
specification with solver queries that check global consistency, vacuousness,
violatability, and redundancy at the requirement level, producing diagnostic
flags that localize structural defects to specific requirements. 
Second, we detect and resolve ambiguity in the source text by exploiting the stochasticity of LLM-based autoformalizers as a diagnostic signal:
sampling multiple independent SMT formalizations of the same requirement and comparing them
with bidirectional SMT equivalence checking; persistent disagreement signals
that the natural-language requirement admits more than one plausible reading.
When defects are detected, we execute a targeted repair through a neurosymbolic feedback loop: solver feedback guides re-formalization for structural defects, and SMT-derived
witnesses guide natural-language clarification for ambiguous requirements. The final requirement is then surfaced to the user for manual edits. 

Our contributions are as follows:
\begin{itemize}[leftmargin=*,noitemsep]
    \item We introduce a neurosymbolic framework for analyzing
    natural-language software requirements. The system, \textsc{VeriMed},
    autoformalizes requirements into a single SMT model and uses the
    solver's output as the primary signal for downstream review.
    
    \item We formulate four requirement-level SMT audits — global
    consistency, vacuousness, violatability, and redundancy — that
    operationalize established requirement-quality criteria. Applied to a
    published hemodialysis specification, the audits flagged 2 of 64
    requirements as redundant.
    
    \item We show that the quality of solver feedback determines the
    effectiveness of LLM repair. In requirement-grounded question
    answering, using the violated requirements as feedback alone raises
    verified accuracy from 55.4\% (no feedback) to 80.0\%; providing an
    SMT counterexample as additional feedback raises accuracy to 98.5\%.
    
    \item We introduce ambiguity detection via bidirectional SMT inequivalence
    across independently sampled formalizations, using solver-generated witnesses
    to drive clarification. To our knowledge,
    this is the first application of solver-checked sample disagreement
    to software requirements analysis. On the hemodialysis benchmark, the
    procedure flagged 12 of 64 requirements (18.8\%) as producing
    multiple distinct encodings; all 12 converged to a single encoding
    after clarification.
\end{itemize}

\section{Methodology}
\subsection{Problem Formulation}

\label{subsec:problem_formulation}
Let $R = \{r_1, \dots, r_n\}$ be a set of natural-language software
requirements. We translate $R$ into a single SMT model $M$ and
use $M$ to answer three classes of queries: requirement-level audits, query
verification against scenario-based safety questions, and ambiguity detection
over the source text.

\textbf{Encoding.} Let $T$ denote the translation from a requirement $r_i$ to
a formal encoding $\rho_i = T(r_i)$. Most safety requirements take a
conditional form: \emph{``if condition $P$ holds, then action $Q$ must
follow.''} We encode each such requirement as
$\rho_i(\mathbf{x}) \equiv p_i(\mathbf{x}) \Rightarrow q_i(\mathbf{x})$,
where $\mathbf{x}$ denotes the typed state variables of the system (Boolean
control states and real-valued device parameters) and $p_i, q_i$ are
quantifier-free formulas over Booleans and linear real arithmetic. Invariant
requirements, those that must hold unconditionally, are encoded directly as
state predicates $\rho_i(\mathbf{x})$ with no antecedent. We also maintain a
set of global domain constraints $C(\mathbf{x})$ encoding background facts
shared across all requirements, such as flow rates being non-negative. The
full model is $M = C \wedge \bigwedge_{i=1}^{n} \rho_i.$

\textbf{Audit queries.} We define four solver queries over $M$ that
characterize the requirement set.
\begin{enumerate}[leftmargin=*,noitemsep]
\item \emph{Global consistency} asks whether $M$ is satisfiable. If $M$ is unsatisfiable, the requirements are mutually inconsistent, and an unsat core identifies a conflicting subset for repair.
\item \emph{Vacuousness} applies to conditional requirements and asks whether the triggering condition $p_i(\mathbf{x})$ can ever hold under the domain
constraints. We check this by asking whether $C \wedge p_i(\mathbf{x})$ is
satisfiable. If it is unsatisfiable, $r_i$ never applies and is therefore
vacuous. Invariant requirements have no antecedent and are not subject to
this check.
\item \emph{Violatability} asks whether a requirement can be broken within the
domain. We check
$C \wedge p_i(\mathbf{x}) \wedge \neg q_i(\mathbf{x})$ for conditional requirements and $C \wedge \neg \rho_i(\mathbf{x})$ for invariants; a satisfiable result returns a concrete violating assignment;
an unsatisfiable result indicates that the violation is already ruled out by
the domain constraints alone.
\item \emph{Redundancy} asks whether $M_{\setminus i} = C \wedge \bigwedge_{j \neq i} \rho_j$ already enforces $r_i$. 
We check $M_{\setminus i} \wedge p_i(\mathbf{x}) \wedge \neg q_i(\mathbf{x})$ for conditional requirements
and $M_{\setminus i} \wedge \neg \rho_i(\mathbf{x})$ for invariants; unsatisfiability in either case
confirms redundancy, and the unsat core identifies the subsuming requirements.
\end{enumerate}

\textbf{Query verification.} Given a scenario-based safety question
translated into a formal claim $\phi$, we ask whether $\phi$ is supported by
the requirements by checking the formula
$M \wedge \neg \phi$. If it is unsatisfiable, $M \models \phi$ and the claim
is entailed; if it is satisfiable, the solver returns a counterexample
showing a state consistent with $M$ but inconsistent with $\phi$, which is then used as feedback for repair.

\textbf{Ambiguity detection.} The previous queries treat $M$ as fixed; we now
consider whether $M$ itself is well-determined by the source requirements. A
natural-language requirement may admit more than one plausible reading, and
the formalizer may silently commit to one. We define ambiguity operationally
through agreement across independently sampled formalizations. Given two candidate encodings $\rho_i^a$ and $\rho_i^b$ of $r_i$, we
say they \emph{agree} if both $C \wedge (\rho_i^a \wedge \neg \rho_i^b)$ and $C \wedge (\rho_i^b \wedge \neg \rho_i^a)$ are unsatisfiable. If the first is satisfiable, the witnessing assignment
exposes a state allowed by $\rho_i^a$ but ruled out by $\rho_i^b$; if the
second is satisfiable, the witness exposes a state allowed by $\rho_i^b$ but
ruled out by $\rho_i^a$. We sample $N$ formalizations of $r_i$ and apply this
pairwise check; a requirement that produces more than one distinct encoding
under bidirectional agreement is flagged as \emph{ambiguity-sensitive}, and
the witnesses guide natural-language clarification. A requirement whose samples all agree is treated as admitting a single formalization under repeated sampling.
\begin{figure}[t]
    \centering
    \includegraphics[width=\linewidth]{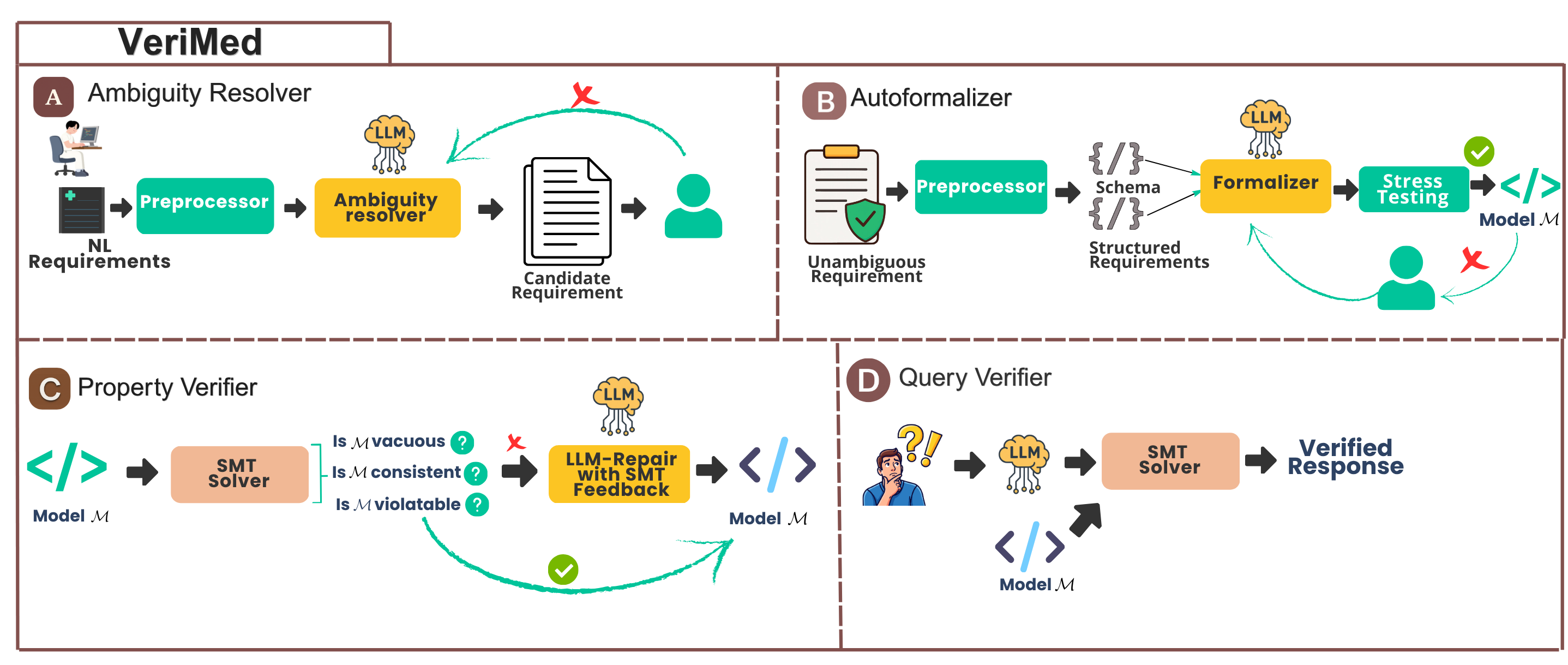}
    \caption{Architecture of \textsc{VeriMed}.}
    \label{fig:method}
\end{figure}
\begin{figure}[t]
    \centering
    \includegraphics[width=\linewidth]{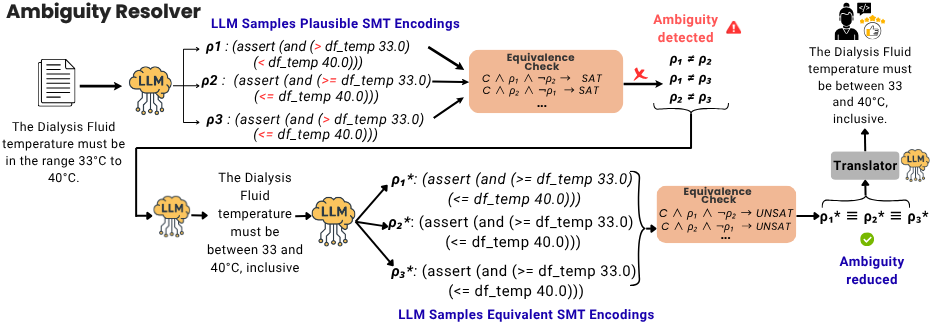}
  \caption{Ambiguity-resolution loop with a dialysate-temperature example.
\textsc{VeriMed} samples $N=5$ independent formalizations and compares them via bidirectional SMT equivalence.
Here, $\rho_1$, $\rho_2$, and $\rho_3$ express different boundary
interpretations of the same temperature range: $\rho_1$ excludes both endpoints,
$\rho_2$ includes both endpoints, and $\rho_3$ excludes 33$^\circ$C but includes
40$^\circ$C. The solver returns concrete disagreement witnesses (e.g., 33$^\circ$C, 40$^\circ$C), which the LLM uses to rewrite the requirement with explicit boundary semantics. The loop repeats until all samples are semantically equivalent, then surfaces the clarified requirement for the user to review.}
\label{fig:ambiguity-resolution-loop}
\end{figure}

\begin{figure}[t]
\centering
\definecolor{accent}{HTML}{1D4ED8}
\definecolor{codefg}{HTML}{0F172A}
\definecolor{rulegray}{HTML}{E2E8F0}
\definecolor{headbg}{HTML}{F8FAFC}
\footnotesize
\setlength{\tabcolsep}{4pt}
\renewcommand{\arraystretch}{0.98}
\begin{tcolorbox}[
  enhanced,
  colback=white,
  colframe=rulegray,
  boxrule=0.4pt,
  arc=2pt,
  left=3pt,right=3pt,
  top=3pt,bottom=3pt,
]
\begin{tabular}{>{\columncolor{headbg}}p{0.44\linewidth}!{\color{black}\vrule width 0.4pt}>{\columncolor{headbg}}p{0.48\linewidth}}
\begin{minipage}[t]{\linewidth}
\vspace{2pt}
{\textcolor{accent}{\textbf{Schema}}}
\vspace{2pt}
\begin{tabular}{@{}lll@{}}
\textbf{Variable} & \textbf{Type} & \textbf{Description} \\
\texttt{bolus\_active} & \textit{Bool} & Bolus active \\
\texttt{volume} & \textit{Real} & Volume (mL) \\
\texttt{alarm\_active} & \textit{Bool} & Alarm active \\
\end{tabular}
\vspace{2pt}
\end{minipage}
&
\begin{minipage}[t]{\linewidth}
\vspace{2pt}\hspace{4pt}%
{\textcolor{accent}{\textbf{NL}}}
\,
\textit{If bolus is active and infused volume exceeds 400\,mL, the alarm shall activate.}
\vspace{6pt}

\hspace{4pt}{\textcolor{accent}{\textbf{SMT}}}
\,
{\ttfamily\scriptsize\textcolor{codefg}{%
(=> (and bolus\_active (> volume 400.0))\\[-1pt]
alarm\_active)}}
\vspace{2pt}
\end{minipage}
\\
\end{tabular}
\end{tcolorbox}
\vspace{-5pt}
\caption{Example variable schema and SMT encoding generated from a requirement.}
\label{fig:autoformalization-example}
\vspace{-8pt}
\end{figure}

\subsection{Approach}\label{sec:approach}
We introduce \textsc{VeriMed}\footnote{Code and data will be released upon acceptance of the paper.}, a neurosymbolic framework for turning
natural-language software requirements into a solver-checkable specification.
\textsc{VeriMed} couples an LLM-based autoformalizer with an SMT solver to run the audit, query verification,
and ambiguity checks defined in Section~\ref{subsec:problem_formulation}.
A central design choice is that all requirements are expressed in the same
symbolic namespace. We define a fixed schema of typed state variables,
covering Boolean control states and real-valued device parameters, together
with the domain constraints $C(\mathbf{x})$. The schema is injected into every
LLM prompt, ensuring independently generated assertions compose into a single model $M$.

\textsc{VeriMed} operates as a four-stage pipeline as shown in Figure ~\ref{fig:method}: the
\textbf{Ambiguity Resolver} screens each requirement for stable
formalization before the requirement is formalized; the
\textbf{Autoformalizer} translates the requirement set into SMT-LIB;
the \textbf{Property Verifier} audits the resulting model; and the
\textbf{Query Verifier} checks proposed answers to scenario-based
safety questions against the model.

\stepbadge{A} \textbf{Ambiguity Resolver.} Natural-language requirements are
prone to ambiguity~\cite{riaz2019automatic}: an LLM autoformalizing an
ambiguous requirement may silently commit to one reading among several. The
Ambiguity Resolver screens each requirement before it enters the formalization
pipeline. For each $r_i \in R$, it samples the LLM $N$ times at elevated
temperature to obtain candidate encodings $\rho_i^1, \dots, \rho_i^N$ and
applies the bidirectional agreement check defined in
Section~\ref{subsec:problem_formulation} to every pair. 
A requirement that produces more than one distinct encoding is flagged as
ambiguity-sensitive: the witnessing assignments from the disagreeing pairs
are surfaced as concrete distinguishing states, and the LLM is prompted to
rewrite the source text until the samples converge. Only requirements that pass screening are
forwarded to the Autoformalizer. Figure ~\ref{fig:ambiguity-resolution-loop} illustrates the clarification loop.

\stepbadge{B} \textbf{Autoformalizer.} The Autoformalizer translates the
disambiguated requirement set into a single SMT-LIB
specification~\cite{barrett2016satisfiability}. The LLM receives the
requirement set together with the canonical schema and the domain constraints,
and produces the requirement predicates $\rho_1, \dots, \rho_n$ along with the
declarations needed to assemble $M$. The output is therefore not a collection
of isolated formulas but a unified model whose assertions can be analyzed
jointly.
If the generated SMT (e.g., Figure \ref{fig:autoformalization-example}) is malformed or rejected by the solver, the system
enters a bounded syntactic repair loop. The solver error message and the
failed output are returned to the LLM as feedback, and regeneration continues
until a syntactically valid model is produced or the retry budget is
exhausted. This loop addresses syntactic faults only; semantic audits are
delegated to the Property Verifier.

\stepbadge{C} \textbf{Property Verifier.} The Property Verifier runs the four
audit queries defined in Section~\ref{subsec:problem_formulation} over the
assembled model $M$: global consistency, vacuousness, violatability, and
redundancy. For each requirement, the verifier records the outcome of every
applicable check and, when an audit fails, returns the diagnostic information
the solver provides: an unsat core for inconsistency or redundancy, and a
concrete violating assignment for violatability. These diagnostics localize
each defect to a specific requirement or subset of requirements and serve as
the input to any downstream repair.

\stepbadge{D} \textbf{Query Verifier.} The Query Verifier tests whether a 
proposed answer is entailed by $M$ under a given scenario. A scenario 
extractor maps the question to a structured state
assignment $s$, an action generator produces a schema-constrained candidate
answer $a$, and the verifier issues the entailment query
$M \wedge s \wedge \neg a$. Unsatisfiability confirms 
entailment by the requirements under the scenario and is accepted; a satisfiable result
returns a counterexample exposing a state consistent
with $M$ and $s$ but inconsistent with $a$. A secondary check identifies the
specific requirements violated in the counterexample, and a response builder
translates the result into a natural-language explanation. When a violation
is found, the repair loop re-prompts the LLM with the violated requirements
and, in the counterexample-guided variant, the solver counterexample itself.

\subsection{Experimental Setup}
\label{sec:experiment}
\textbf{Domain and benchmark.}
We evaluate \textsc{VeriMed} on the hemodialysis machine case study of
Mashkoor~\citep{mashkoor2016hemodialysis}, using a corpus of 64
natural-language safety requirements drawn from the software requirements
specification. The requirements cover alarm conditions, flow constraints,
connectivity constraints, and numeric limits across multiple machine phases
and subsystems.

\textbf{Schema, formalism, and solver.}
All experiments share the same canonical typed schema and the same global
domain constraints $C(\mathbf{x})$. The schema contains 85 typed variables
covering Boolean machine states and real-valued device parameters. All
formalizations are expressed in SMT-LIB2 over quantifier-free linear real
arithmetic with Booleans. We use Z3~\cite{de2008z3} for all SMT checks.

\textbf{Language models.}
The primary LLM is \texttt{claude-sonnet-4-6}, used for SMT generation,
answer generation, and repeated formalization across all experiments. Temperature is 0 for deterministic outputs (answer
generation and repair), 1.0 for ambiguity detection (exposing alternative encodings), 
and 0.2 for clarification of ambiguous requirements. A secondary model, \texttt{claude-haiku-4-5-20251001}, is used at temperature 0 for the
lightweight answer-to-assignment translation step, with a rule-based
fallback parser if translation fails. SMT generation uses a 16{,}384-token
output budget and a syntactic repair loop of up to five retries. LLM calls
are stateless: each call reconstructs the full prompt, with explicit repair
feedback appended when applicable.

\subsection{Experiment 1: Requirement Autoformalization}
Before any audit can be trusted, the formalization must be trustworthy. This experiment
evaluates whether \textsc{VeriMed} can translate the 64 hemodialysis safety
requirements into a single solver-checkable SMT model that is semantically faithful to the source, covering
generation quality, round-trip fidelity, and sensitivity to
requirement-level faults.

\textbf{Setup: Generation.} The Autoformalizer receives the full requirement set
together with the canonical typed schema and produces the complete SMT-LIB
model in a single batch translation, rather than encoding each requirement
independently. If the generated
output fails to parse or is rejected by Z3, the autoformalizer regenerates
with the solver's error message as feedback, capped at five retries. We then
run the four audit queries from Section~\ref{subsec:problem_formulation} over
the assembled model.

\textbf{Round-trip semantic equivalence check.} To test whether the generated SMT
is semantically faithful to the source requirements, following the round-trip
faithfulness check approach of Amrollahi et al.~\cite{amrollahi2026faithful}, we run a round-trip translation
cycle. The generated model is first translated back into structured
natural-language requirements which are then
re-formalized under the same canonical schema. Each reconstructed
formula is compared to its original under bidirectional SMT equivalence. We
also report the mean cosine similarity between original and reconstructed
requirement texts using normalized embeddings from \texttt{all-mpnet-base-v2}.

\textbf{Mutation stress test.} To test model sensitivity
to requirement-level faults, we construct 256 mutated requirements
spanning four fault categories drawn from FDA MAUDE adverse-event reports for
hemodialysis machines~\cite{fdamaude}: \textit{false-alarm trigger
suppression}, \textit{mode-transition guard removal}, \textit{value mismatch},
and \textit{limit violation}. Each mutated requirement is compared against
the canonical model using bidirectional SMT equivalence; a mutation is
detected if either direction returns \textsc{sat}.

\textbf{Metrics.} We report parse success rate, the four requirement-level
audit outcomes, round-trip equivalence and mean cosine similarity, and per-category and
overall mutation detection rates. Binomial rates are reported with two-sided
Wilson 95\% confidence intervals.

\textbf{Result and Insight.} The results are shown in Table~\ref{tab:exp1_audit}. 
The Autoformalizer produced a solver-checkable SMT model
covering all 64 requirements on the first generation attempt. The resulting
model was globally consistent and contained no vacuous requirements; 2 
requirements were redundant under the rest of the specification. All 64
requirements survived the round-trip check: every reconstructed requirement
re-formalized to a bidirectionally equivalent formula, 
with mean cosine similarity of 96.2\% between original and reconstructed texts. 
Under mutation testing, the audit pipeline
detected 249 of 256 single-requirement mutations
(97.3\%, Wilson 95\% CI [94.5\%, 98.7\%]) across the four fault categories.
\textbf{Observation: Natural-language safety requirements contain latent
defects that solver-level checks make visible.}
Our approach flagged $r_{11}$ and $r_{56}$ as redundant. For $r_{11}$, the unsat core
contains $r_{55}$: $r_{11}$ concerns venous pressure falling below its
lower limit, while $r_{55}$ requires venous pressure to remain at least
22.5 mmHg above that limit, making $r_{11}$'s trigger unreachable. 
For $r_{56}$, the unsat core contains $r_{42}$: $r_{56}$ concerns
dialysate conductivity outside the permissible range, while $r_{42}$
requires conductivity to stay between 12.5 and 16.0 mS/cm. 
These requirements are not locally vacuous under the domain constraints
alone; their guards are satisfiable against $C$. They become redundant
only once the surrounding requirements are added. This makes redundancy a
review signal rather than a verdict: the engineer should inspect whether
the subsuming requirement is intentionally ruling out the condition, or
whether one of the requirements is too strong. Details anaylysis on the redundancy is provided in Appendix \ref{app:redundancy}. 

\textbf{Generalizability.} To test whether
the round-trip results transfer beyond the hemodialysis benchmark,
we applied the same pipeline end-to-end to an open-source PCA
(patient-controlled analgesia) pump specification with 144
requirements: the autoformalizer produced an SMT model from the new
requirement set, and the round-trip check ran on that model. The
cycle reconstructed all 144 requirements; initial bidirectional SMT
equivalence held for 135 of 144 (93.75\%), and a single repair
iteration closed the remaining 9 cases, reaching full equivalence
(144/144). Mean cosine similarity between original and reconstructed
requirement texts was 0.87, lower than the hemodialysis result, with
the largest textual gaps concentrated in compliance-related
requirements where reconstruction paraphrased more aggressively.
Despite the lexical drift, the repaired re-formalizations recovered
the original SMT semantics exactly. Details of the audited redundant requirements are in
Appendix~\ref{app:redundancy}.
\begin{table}[t]
\centering
\caption{Experiment 1 results. \textbf{(a)} Audit and round-trip results on
64 requirements. \textbf{(b)} Mutation stress test on 256 mutated cases.
RT = round-trip; CIs are Wilson 95\% for binomial rates.}
\label{tab:exp1_audit}
\setlength{\tabcolsep}{5pt}
\renewcommand{\arraystretch}{1.13}

\definecolor{hdrblue}{HTML}{1E3A5F}
\definecolor{rowgray}{HTML}{F3F6FA}

\begin{minipage}[t]{0.48\columnwidth}
\centering
\scriptsize
\rowcolors{2}{rowgray}{white}
\begin{tabular}{lccc}
\rowcolor{hdrblue}
\textcolor{white}{\textbf{Metric}} &
\textcolor{white}{\textbf{Frac.}} &
\textcolor{white}{\textbf{\%}} &
\textcolor{white}{\textbf{95\% CI}} \\
Requirements covered     & $64/64$ & $100.0$ & $[94.3, 100.0]$ \\
Non-vacuous (cond.)      & $39/39$ & $100.0$ & $[91.0, 100.0]$ \\
Violatable               & $64/64$ & $100.0$ & $[94.3, 100.0]$ \\
Redundant                &  $2/64$ &   $3.1$ & $[0.9,  10.7]$  \\
RT equivalence           & $64/64$ & $100.0$ & $[94.3, 100.0]$ \\
RT cosine sim.\ (mean)   &   ---   &  $96.2$ & ---             \\
\end{tabular}
\vspace{4pt}

{\scriptsize\textit{(a) Audit and round-trip.}}
\end{minipage}
\hfill
\begin{minipage}[t]{0.48\columnwidth}
\centering
\scriptsize
\rowcolors{2}{rowgray}{white}
\begin{tabular}{lccc}
\rowcolor{hdrblue}
\textcolor{white}{\textbf{Mutation type}} &
\textcolor{white}{\textbf{Frac.}} &
\textcolor{white}{\textbf{\%}} &
\textcolor{white}{\textbf{95\% CI}} \\
False alarm      & $64/64$   & $100.0$ & $[94.3, 100.0]$ \\
Mode transition  & $63/64$   &  $98.4$ & $[91.7,  99.7]$ \\
Value mismatch   & $62/64$   &  $96.9$ & $[89.3,  99.1]$ \\
Limit violation  & $60/64$   &  $93.8$ & $[85.0,  97.5]$ \\
\rowcolor{hdrblue!10}
\textbf{Overall} & $249/256$ &  $97.3$ & $[94.5,  98.7]$ \\
\end{tabular}
\vspace{4pt}

{\scriptsize\textit{(b) Mutation stress test.}}
\end{minipage}
\end{table}

\subsection{Experiment 2: Counterexample-Guided Repair (CEGR)}
We evaluate whether SMT feedback repairs incorrect LLM answers about
mandatory machine actions, and in particular whether concrete counterexamples
drive more repair than symbolic feedback alone.

\textbf{Setup.} We evaluate on 65 scenario-based safety questions derived
from the HD machine case study, drafted with LLM assistance and
reviewed manually to ensure that each question maps to a specific requirement
subset and a solver-verifiable answer. Each question describes a concrete
machine state and asks what mandatory device actions follow. The benchmark
contains \textit{single-requirement} queries (29), \textit{multi-requirement} 
queries (15), and \textit{no-requirement} queries (21). For each
question, the LLM produces a structured JSON answer constrained to schema
variables relevant to that question. Given model $M$, scenario $s$, and candidate answer $a$, the verifier
checks $M \wedge s \wedge \neg a$, labeling results \textsc{safe} (\textsc{unsat}, no counterexample exists), \textsc{violation} (\textsc{sat}, a counterexample exists), or \textsc{unknown}. Only
\textsc{safe} answers are counted as correct. All conditions use the same prompt,
answer format, and verifier, differing only in rejection feedback: \textit{baseline} (no feedback),
\textit{self-repair} (generic rejection message), \textit{requirement feedback only}
(violated requirements only), full \textit{CEGR} (violated requirements and Z3 counterexample witness).
Each condition has one initial attempt and up to five repair iterations.

\textbf{Metrics.} We report first-attempt accuracy (pass@1), final accuracy
after the repair budget, and repair success rate. Accuracy is the fraction of
questions labeled \textsc{safe} by the verifier. Repair success rate is the
fraction of initially non-\textsc{safe} answers that become \textsc{safe}
within the repair budget.

\begin{figure}[t]
    \centering
    \includegraphics[width=0.99\linewidth]{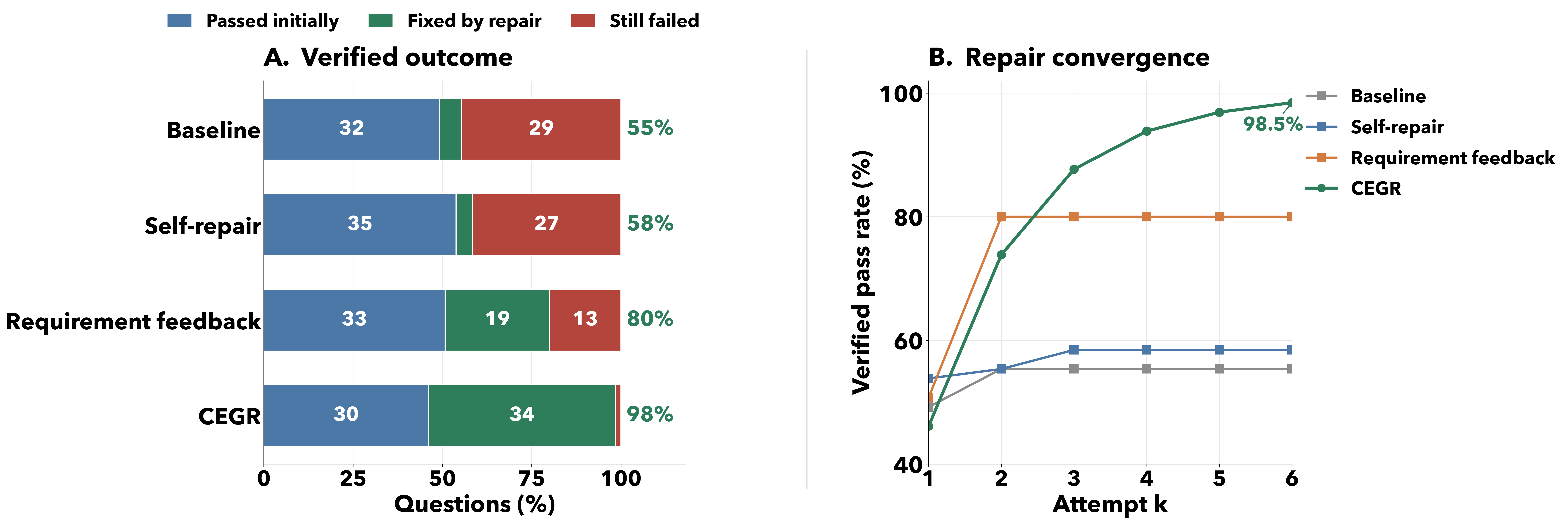}
    \caption{Effect of verifier feedback on question answering. (Left) accepted on the first attempt, fixed during repair, and rejected after the repair budget. (Right) Cumulative SMT-verified accuracy.}
\label{fig:exp2-cegr-main}
    \label{fig:exp2}
\end{figure}

\textbf{Result and Insight.} Figure~\ref{fig:exp2-cegr-main} shows the results. CEGR achieves 98.5\% final accuracy, exceeding the no-feedback baseline (55.4\%), self-repair (58.5\%), and requirement-only feedback (80.0\%). First-attempt accuracy is comparable across conditions (baseline 49.2\%, self-repair 53.8\%, requirement-only 46.2\%, CEGR 46.2\%); the divergence emerges entirely from repair behavior. Repair success rate follows the same pattern: 12.1\%, 10.0\%, 59.4\%, and 97.1\%, respectively. The 18.5-point gap between requirement-only feedback and full CEGR shows that the counterexample, not solver feedback in general, is what drives repair: returning a concrete witness state lets the LLM localize the fault in its previous answer in a way that violated requirements alone do not.
\textbf{Observation: With the LLM and prompt held fixed, the choice of
feedback signal impacts repair outcomes.} First-attempt accuracy across the four conditions ranged narrowly from
46.2\% to 53.8\%. After up to five repair iterations, final accuracy ranged
from 55.4\% to 98.5\%, and repair success rate from 10.0\% to 97.1\%. The
conditions differed only in what the verifier returned after a rejected
answer, so the spread is attributable to the feedback signal.
See Appendix \ref{tab:exp2_question_sample} for a sample of questions and verified responses.

\subsection{Experiment 3: End-to-End Fault Detection}
Experiment 1 showed that the canonical SMT model distinguishes mutated formulas 
under bidirectional equivalence.
This experiment tests the complementary claim: whether the full pipeline,
including re-autoformalization, catches faults injected one at a time into the
requirement set.

\textbf{Setup.} For each of the 64 hemodialysis requirements $r_i$, we manually
write a mutated natural-language version $r_i^{\text{mut}}$ representing a
plausible safety defect: a changed threshold, a weakened trigger condition, or
an incorrect control action. We autoformalize each mutated requirement into a
mutant SMT block $\rho_i^{\text{mut}}$, and construct the modified model
$M^{(i)}$ by replacing $\rho_i^{\text{original}}$ with $\rho_i^{\text{mut}}$ in
the canonical model while keeping the shared schema, the global domain
constraints $C$, and the remaining 63 requirement blocks fixed.

\begin{figure}[t]
    \centering
    \includegraphics[width=0.8\columnwidth]{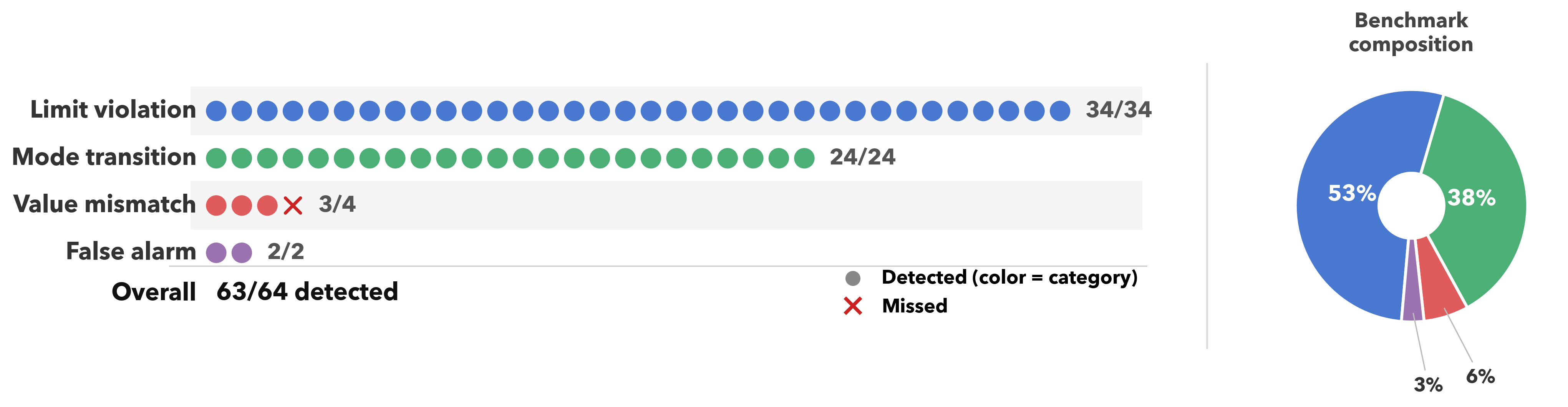}
   \caption{Mutation Detection. (Left) detection outcomes
by fault type: circles denote a detected mutant, a red cross denotes
the single missed mutant. (Right) benchmark
composition by fault type.}
    \label{fig:exp5}
\end{figure}

\textbf{Evaluation.} A fault is detected if either of two conditions holds:
(a) the modified model $M^{(i)}$ is globally inconsistent, indicating that
$r_i^{\text{mut}}$ contradicts the original requirements; or (b) the mutant
and original requirements are semantically distinguishable, checked by
bidirectional satisfiability between $\rho_i^{\text{mut}}$ and
$\rho_i^{\text{original}}$ under $C$. We test whether
$C \wedge \rho_i^{\text{mut}} \wedge \neg \rho_i^{\text{original}}$ is
satisfiable (the mutant permits a behavior ruled out by the original
requirement) and whether
$C \wedge \rho_i^{\text{original}} \wedge \neg \rho_i^{\text{mut}}$ is
satisfiable (the original requirement allows a behavior ruled out by the
mutant); a satisfiable result in either direction signals detection. This
disjunction separates faults that produce immediate contradictions from faults
that remain satisfiable but still change the intended safety behavior.

\textbf{Result and Insight.} Figure ~\ref{fig:exp5} shows the complete result: \textsc{VeriMed} detected 63 of 64 inserted defects
(98.4\%, Wilson 95\% CI [91.7\%, 99.7\%]). All 64 mutated requirement sets
autoformalized successfully and remained globally satisfiable, 
so detection came entirely from comparing original and mutated encodings. Detection was complete
for limit violations (34/34), mode-transition changes (24/24), and
false-alarm changes (2/2). For value mismatches, 3 of 4 mutants were
detected; the missed case was the mutant for $r_2$, which was
indistinguishable from the original under the current schema. 
\textbf{Observation: Consistency alone does not always catch requirement faults}. 
Every mutated specification remained
globally satisfiable, yet 63 of 64 defects were detected. 
The $r_{20}$ mutant shows why: the original requires disconnection and an alarm when dialysate temperature falls below 33$^\circ$C; the mutant requires the opposite.
Both are  still globally
satisfiable, but only the original reflects the intended safe behavior. Internal consistency alone is therefor insufficient; the useful signal is behavioral preservation.

\subsection{Experiment 4: Ambiguity Resolution}
We evaluate whether the SMT-solver feedback can be used not only to detect ambiguous requirements, but also to reduce ambiguity in the natural-language specification. 

\begin{figure}[t]
    \centering
    \includegraphics[width=\columnwidth]{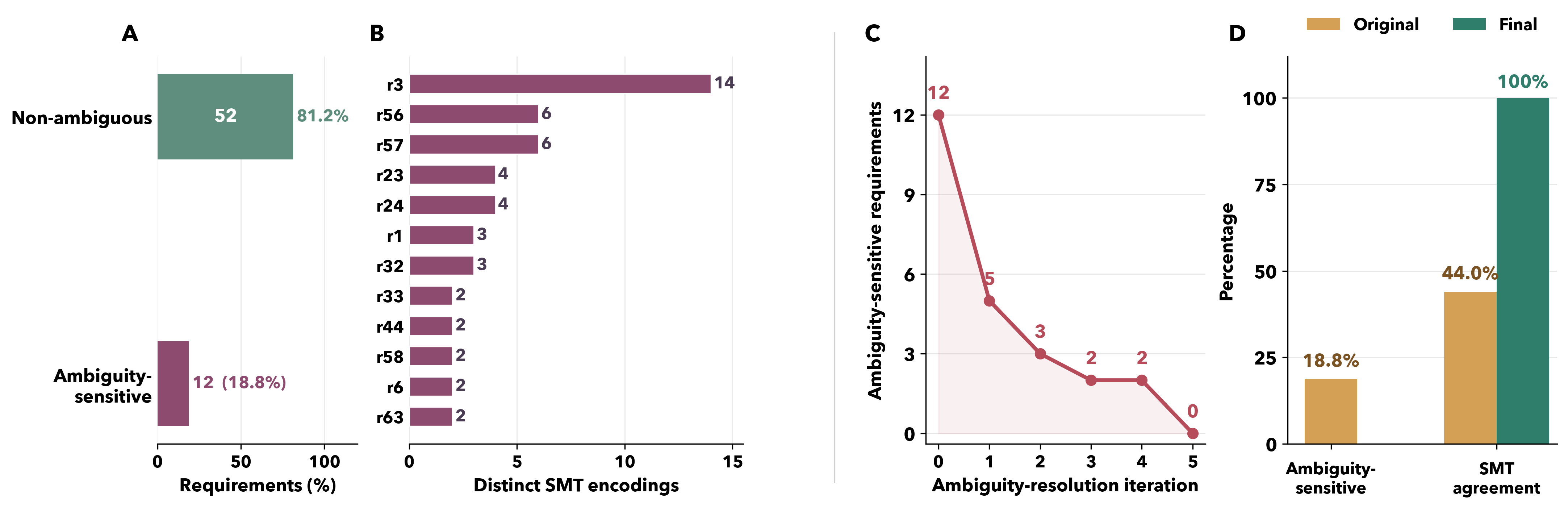}
    \caption{Ambiguity detection and resolution via repeated autoformalization. (A) Share of requirements flagged as ambiguity-sensitive versus non-ambiguous. (B) Distinct SMT encodings per flagged requirement, ranked by count. (C) Ambiguity-sensitive count across five rounds of SMT-guided clarification. (D) Pairwise SMT agreement.}
    \label{fig:exp6}    
\end{figure}

\textbf{Setup}: We apply the Ambiguity Resolver (Section~\ref{sec:approach}) to all 64 hemodialysis requirements with $N = 5$ samples per requirement. Requirements that remain ambiguity-sensitive after clarification are re-screened; the clarified requirements are stitched back into the full 64-requirement set.
\textbf{Result and Insight}: Figure~\ref{fig:exp6} shows the results.
12 out of the 64 requirements were flagged as ambiguity-sensitive (18.8\%, CI [11.1\%, 30.0\%]); pairwise SMT agreement across all formalization pairs is 44.0\% (CI [41.7\%, 46.3\%]). After SMT-guided clarification, ambiguity-sensitive requirements drop to 0/64 (0.0\%, CI [0.0\%, 5.7\%]) and pairwise SMT agreement reaches 100\% (CI [99.4\%, 100.0\%]), with mean distinct encodings per requirement falling from 1.59 to 1.00. All 320 clarified formalizations remain syntactically valid and locally verified, confirming that the clarification loop eliminates encoding ambiguity without introducing inconsistency.
\textbf{Observation: SMT disagreement across samples flags requirements that the formalizer reads in more than one way.} One recurring
source of ambiguity is overlapping numeric ranges: R-24 and R-25 \cite{mashkoor2016hemodialysis} specify
air-detection thresholds for \emph{``0 to 200 mL/min''} and
\emph{``200 to 400 mL/min,''} leaving the endpoint 200 ambiguous. The autoformalizer split
on this, producing encodings that disagreed at exactly 200 mL/min.
Solver-generated witnesses surfaced each such disagreement as a concrete
state, and the clarification loop converged on a single reading for every
flagged requirement (see Appendix~\ref{tab:exp6_original_vs_clarified_requirements} for the full list).

\section{Limitations}
\textbf{Generalizability.} Our evaluation targets hemodialysis machine
requirements, with a secondary evaluation on a PCA pump specification
(Appendix~\ref{app:pca-generalizability}). Both are infusion-related
medical devices with shared structural features. We cannot establish
that our findings transfer to structurally different device classes
or to non-medical domains, and we make no such claim.

\textbf{Faithfulness.} Our approach does not directly measure whether an LLM-generated formalization is fully faithful to the intent of the original requirement. That is a non-trivial challenge. 
Instead, we use a collection of proxy signals, including consistency,
mutation sensitivity, ambiguity detection, round-trip similarity, and
downstream repair performance. These signals are useful but incomplete: a
formalization can pass some of them while still missing subtle aspects of the
intended meaning. 

\textbf{Temporal Requirements. }A further limitation of our work is temporal expressiveness. Since we use SMT as our core formalism, temporal requirements are modeled using current
state conditions rather than a full temporal logic. For example, requirements
such as ``the blood pump stops and remains stopped'' or ``after 400\,mL have
been reinfused, the alarm shall activate'' are approximated in terms of the
current system state instead of temporal relations across states.

\textbf{Scalability and Deployment Risk.} Industrial medical software requirements documents are often an order of
magnitude larger than the datasets used in this work. The result of this work should not be used in safety-critical
applications without independent review by experts.

\section{Related Work}
\textbf{Autoformalization}
Since \citep{wu2022autoformalization} demonstrated LLM-based autoformalization for mathematical theorems, neurosymbolic research has focused on combining autoformalization with downstream verification~\citep{bayless2025neurosymbolic, liu2026multi, wu2022autoformalization, akinfaderin2025verafi}. However, the autoformalization step itself is rarely scrutinized. In particular, existing work rarely evaluates whether the generated formalizations are consistent, correct, and unambiguous as specifications \citep{sun2024clover, yang2023leandojo, jiang2022draft}. Prior work on autoformalization has largely focused on mathematical domains, where formal statements are well-defined and can be verified against proofs \cite{allamanis2024unsupervised, li2024autoformalize, chen2025reform}. However, \textsc{VeriMed} operates on natural language software requirement specifications: theorems are self-contained and unambiguous by construction, while natural-language requirements often contain underspecification and ambiguity that the formalization step must surface rather than ignore. \textsc{VeriMed} addresses both gaps: it checks the correctness of the formalization itself, and uses verifier feedback to drive repair.

\textbf{Ambiguity Resolution.} Ambiguity is a central obstacle in formalizing natural-language requirements. Recent work uses LLMs to detect and explain ambiguities~\citep{bashir2025requirements}, or surfaces them directly during code generation via clarifying questions~\citep{mu2024clarifygpt, wu2025can, wu2025humanevalcomm}. These approaches treat ambiguity as a textual property to flag or paraphrase away without deterministic resolution. Our approach forces ambiguities to manifest in the form of multiple machine-checkable inequivalent encodings of the same requirement. 

\textbf{Requirements Formalization with LLMs.}
Recent efforts evaluate correctness only at the artifact level. \citet{cao2025informal} benchmark ten LLMs across five proof targets, scoring outputs by whether the checker accepts the output; however, acceptance can conceal incorrect properties. \citet{mandal2023large} synthesize configuration specifications from software manuals, but evaluate only against gold labels with no semantic check. \citet{beg2025formalising} survey LLM-based specification tools and frame the open problem as bridging informal natural language and rigorous formal specifications, with no system closing the loop at the requirement level. \textsc{VeriMed} closes this loop, using SMT equivalence between the original encoding and a re-formalization of the LLM's own informalization to detect semantic drift at the requirement level.

\textbf{Formal Analysis of Medical Device Software Requirements.}
Medical device software has long served as a testbed for requirements formalization. The FDA Generic Patient-Controlled Analgesia infusion pump was formalized as timed automata in UPPAAL by \citet{kim2011safety}, and the hemodialysis machine case study introduced at ABZ 2016 \citep{mashkoor2016hemodialysis} has since been formalized in Circus \citep{gomes2016modelling}, Event-B \citep{hoang2016validating}, and Abstract State Machines \citep{arcaini2016assure}, each constructed manually by formal-methods experts. More recent LLM-based work has begun to automate this translation: \citet{wang2025supporting} pair Claude 3.5 Sonnet with the ESBMC bounded model checker on nine Lockheed Martin cyber-physical systems.

\section{Conclusion and Future Work}

We presented \textsc{VeriMed}, a neurosymbolic framework for autoformalizing
and auditing natural-language medical-device software requirements with SMT
solver feedback. Because software requirements typically lack a gold formal
target, \textsc{VeriMed} relies on two solver-grounded signals: satisfiability
queries over the assembled formal model and agreement among independently
sampled formalizations. We show that concrete SMT counterexamples are the
strongest repair signal for requirement-grounded question answering, and that
repeated formalization exposes hidden ambiguity. Future work should extend
\textsc{VeriMed} to automated schema construction and temporal formalisms.

\bibliography{biblography}

\appendix
\newpage
\clearpage
\section{Generalizability Check: PCA Pump Requirements}
\label{app:pca-generalizability}

To assess whether the round-trip faithfulness evaluation generalizes beyond the
hemodialysis benchmark, we applied the same evaluator to a second medical device
domain: a PCA pump requirement set~\cite{pcapump}. The setup mirrors Experiment 1, we used the same LLM
and a similar repair budget of 5 iterations was applied to 144 PCA pump requirements formalized into a separate SMT model.

\begin{table}[t]
\centering
\caption{Round-trip evaluation on the PCA pump requirement set. Formal
equivalence is measured between the original SMT predicates and those
obtained by reconstructing requirements from the model and re-formalizing.}
\label{tab:pca-roundtrip}
\setlength{\tabcolsep}{5pt}
\renewcommand{\arraystretch}{1.13}
\scriptsize

\definecolor{hdrblue}{HTML}{1E3A5F}
\definecolor{rowgray}{HTML}{F3F6FA}

\rowcolors{2}{rowgray}{white}
\begin{tabular}{lc}
\rowcolor{hdrblue}
\textcolor{white}{\textbf{Metric}} &
\textcolor{white}{\textbf{Value}} \\
Requirements reconstructed            & 144 / 144 \\
Missing requirements                  & 0 \\
Mean cosine similarity                & 0.8688 \\
Median cosine similarity              & 0.8868 \\
90th percentile cosine similarity     & 0.9688 \\
Initial formal equivalence            & 135 / 144\ \ (93.75\%) \\
Final formal equivalence after repair & 144 / 144\ \ (100.0\%) \\
Non-equivalent requirements repaired  & 9 / 9 \\
Repair attempts needed                & 1 \\
\end{tabular}
\end{table}

Textual similarity was high overall: median cosine similarity 0.8868,
90th percentile 0.9688. The lowest-similarity cases were concentrated in
standards, compliance, and security requirements, where reconstruction
paraphrased the source text more aggressively. At the SMT level, 135 of 144
requirements were equivalent on the initial re-formalization pass; the bounded
repair loop resolved all nine remaining mismatches in a single attempt.
Common mismatch patterns included enum and value remapping
errors in alarm and infusion-mode encodings, dropped \texttt{else} branches in
conditional behavior, omitted structural constraints from table-driven
requirements, and symbolic-versus-literal substitutions in alarm-table ranges.

These results mirror the HD round-trip findings: textual reconstruction may
paraphrase the original requirement, yet the SMT-level check can still recover
the original formal semantics. After bounded repair, every reconstructed PCA
requirement formalized back to a predicate equivalent to the original in the
generated model. This provides evidence that the round-trip evaluator is not
specific to the HD benchmark and can serve as an internal consistency check for
autoformalized requirement models across medical-device domains.
\section{Redundancy Analysis}
\label{app:redundancy}

\begin{table}[h]
\centering
\caption{The two requirements flagged as redundant by \textsc{VeriMed}'s
audit ($r_{11}$, $r_{56}$), shown alongside the surrounding requirements
that subsume them ($r_{55}$, $r_{42}$). Each subsuming requirement
appears in the unsat core of the corresponding redundancy check.}
\label{tab:redundancy_examples}
\setlength{\tabcolsep}{6pt}
\renewcommand{\arraystretch}{1.25}
\small

\definecolor{hdrblue}{HTML}{1E3A5F}
\definecolor{rowgray}{HTML}{F3F6FA}

\rowcolors{2}{rowgray}{white}
\begin{tabular}{cp{0.78\linewidth}}
\rowcolor{hdrblue}
\textcolor{white}{\textbf{ID}} &
\textcolor{white}{\textbf{Requirement}} \\
$r_{11}$ & During initiation, if the pressure at the VP transducer
falls below the lower pressure limit, the software shall stop the BP
and execute an alarm signal. \\
$r_{55}$ & The minimum distance between the venous lower pressure
limit and the actual VP is always at least 22.5 mmHg. \\
$r_{56}$ & Bypass mode occurs when the DF conductivity goes beyond
permissible limits; the dialyzer is separated from the DF flow. \\
$r_{42}$ & The DF conductivity parameter must be within 12.5 to
16.0 mS/cm. \\
\end{tabular}
\end{table}

\begin{table}[t]
\centering
\caption{Experimental setup summary. All SMT queries use Z3.}
\label{tab:setup}
\setlength{\tabcolsep}{5pt}
\renewcommand{\arraystretch}{1.13}
\scriptsize

\definecolor{hdrblue}{HTML}{1E3A5F}
\definecolor{rowgray}{HTML}{F3F6FA}

\rowcolors{2}{rowgray}{white}
\begin{tabular}{llclcc}
\rowcolor{hdrblue}
\textcolor{white}{\textbf{Exp}} &
\textcolor{white}{\textbf{Task}} &
\textcolor{white}{\textbf{$N$}} &
\textcolor{white}{\textbf{Model}} &
\textcolor{white}{\textbf{Temp}} &
\textcolor{white}{\textbf{Iter.\ / Samples}} \\
Exp1 & Formalization audit           &  64  & Sonnet 4.6                        & default          & 5 repairs     \\
Exp2 & Question answering \& repair  &  65  & Sonnet 4.6 / Haiku 4.5$^\dagger$  & 0                & 5 iterations  \\    
Exp3 & End-to-end fault detection    &  64  & Sonnet 4.6                        & default          & 6 repairs     \\
Exp4 & Ambiguity detection           &  64  & Sonnet 4.6                        & $1.0\,/\,0.2^\ddagger$ & $N{=}5$ samples \\
\end{tabular}

    \vspace{4pt}
\raggedright
$^\dagger$ Haiku 4.5 (temp.\,0) for answer-to-schema translation; Sonnet 4.6 for generation and repair.\\
$^\ddagger$ Temp.\,1.0 for repeated formalization; 0.2 for SMT-guided clarification.
\end{table}
\newpage

\section{Tables}

\begin{table*}[t]
\centering
\caption{Original and clarified natural-language requirements for the 12 requirements whose repeated formalizations were ambiguity-sensitive before SMT-guided clarification.}
\label{tab:exp6_original_vs_clarified_requirements}
\setlength{\tabcolsep}{3.5pt}
\renewcommand{\arraystretch}{1.08}
\scriptsize
\begin{tabular}{lcclp{0.39\linewidth}p{0.39\linewidth}}
\toprule
\textbf{Req.} & \textbf{Before} & \textbf{Final} & \textbf{$\Delta$} & \textbf{Original requirement} & \textbf{Clarified requirement} \\
\midrule
\rowcolor{purple!8}
r3 & 14 & 1 & -13 & To prevent coagulation of blood, the anti-coagulation pump doses anti-coagulant into the bloodline between the BP and the dialyzer at the set rate or set volume during treatment. & To prevent coagulation of blood, the anti-coagulation pump doses anti-coagulant into the bloodline between the BP and the dialyzer at exactly the set rate during treatment, while the patient is connected and anti-coagulant delivery is active and not stopped. \\
\rowcolor{purple!8}
r56 & 6 & 1 & -5 & Bypass mode occurs when the DF conductivity goes beyond permissible limits; the dialyzer is separated from the DF flow. & Whenever the DF conductivity goes outside permissible limits (below 12.5 mS/cm or above 16.0 mS/cm), both of the following consequences must hold simultaneously: (1) bypass mode becomes active, and (2) the dialyzer is disconnected from the DF flow. Both consequences are required together under this single trigger condition; neither consequence is optional or independent of the other. \\
\rowcolor{purple!8}
r57 & 6 & 1 & -5 & Bypass mode occurs when the DF temperature goes beyond permissible limits; the dialyzer is separated from the DF flow. & If the DF temperature is outside permissible limits (below 33°C or above 40°C), then bypass mode is active and the dialyzer is disconnected from the DF flow. \\
\rowcolor{orange!7}
r23 & 4 & 1 & -3 & If the flow through the SAD sensor is in the range of 0 to 200 mL/min, then the air volume limit for air detection shall be 0.2 mL. & If the flow through the SAD sensor is greater than or equal to 0 mL/min and less than or equal to 200 mL/min, then the air volume limit for air detection shall be 0.2 mL. \\
\rowcolor{orange!7}
r24 & 4 & 1 & -3 & If the flow through the SAD sensor is in the range of 200 to 400 mL/min, then the air volume limit for air detection shall be 0.3 mL. & If the flow through the SAD sensor is greater than 200 mL/min and less than or equal to 400 mL/min, then the air volume limit for air detection shall be 0.3 mL. \\
\rowcolor{orange!7}
r1 & 3 & 1 & -2 & Arterial and venous connectors of the EBC are connected to the patient simultaneously. & The arterial connector of the EBC is connected to the patient if and only if the venous connector of the EBC is also connected to the patient. \\
\rowcolor{orange!7}
r32 & 3 & 1 & -2 & If the machine is in the initiation phase and net fluid removal is enabled and backward rotation of the UFP is detected, the software shall put the machine in bypass and execute an alarm signal. The backward delivered volume shall not exceed 400 mL. & If the machine is in the initiation phase and net fluid removal is enabled and backward rotation of the UFP is detected, the software shall simultaneously put the machine in bypass mode, activate an alarm signal, and ensure the backward delivered volume does not exceed 400 mL; all three responses are required together under these three preconditions. \\
\rowcolor{gray!7}
r33 & 2 & 1 & -1 & If the machine is in the initiation phase and net fluid removal is enabled and the net fluid removal volume exceeds (UF set volume + 200 mL), the software shall put the machine in bypass and execute an alarm signal. & If the machine is in the initiation phase and net fluid removal is enabled and the net fluid removal volume exceeds (UF set volume + 200 mL), the software shall activate bypass mode and execute an alarm signal. \\
\rowcolor{gray!7}
r44 & 2 & 1 & -1 & The DF temperature parameter must be within 33 to 40°C. & The DF temperature setting must be within 33 to 40°C inclusive. \\
\rowcolor{gray!7}
r58 & 2 & 1 & -1 & If the DF concentration is incorrect, the dialyzer will be bypassed. & If the DF concentration is incorrect, bypass mode shall be activated and the dialyzer shall be disconnected from the DF flow. \\
\rowcolor{gray!7}
r6 & 2 & 1 & -1 & Once empty dialyzer has been confirmed, the BP cannot be started anymore. & Once empty dialyzer has been confirmed, the blood pump (BP) cannot be started or kept running anymore. \\
\rowcolor{gray!7}
r63 & 2 & 1 & -1 & During reinfusion, the BP stops automatically after 400 mL have been reinfused. & During reinfusion, the blood pump stops automatically and remains stopped once 400 mL or more have been reinfused, meaning the blood pump is not running. \\
\bottomrule
\end{tabular}
\end{table*}

\begin{table*}[t]
\centering
\caption{Sample of 10 handcrafted Experiment~2 evaluation questions, covering all three query types.}
\label{tab:exp2_question_sample}
\small
\setlength{\tabcolsep}{6pt}
\renewcommand{\arraystretch}{1.15}
\begin{tabular}{p{0.72\textwidth}p{0.22\textwidth}}
\toprule
\textbf{Question} & \textbf{Gold answer} \\
\midrule
\midrule
\multicolumn{2}{l}{\textbf{Single-Requirement Queries}} \\
During initiation, actual blood flow is 90 mL/min, target blood flow is 200 mL/min, and bypass mode is inactive. What are all the mandatory machine actions? If none are required, answer: "No mandatory machine action." & Trigger an alarm. \\
During initiation, net fluid removal is enabled and the UFP is rotating backward. What are all mandatory machine actions? If none are required, answer: "No mandatory machine action." & Activate bypass mode and trigger an alarm. \\
In this machine state, the patient's blood pressure has dropped. What are all mandatory machine actions? If none are required, answer: "No mandatory machine action." & Activate saline infusion and close arterial access. \\
During initiation, DF temperature is 43 C. What are all mandatory machine actions? If none are required, answer: "No mandatory machine action." & Disconnect the dialyzer from df flow, activate bypass mode, and trigger an alarm. \\
\midrule
\multicolumn{2}{l}{\textbf{Multi-Requirement Queries}} \\
During initiation, anticoagulant delivery is running, reverse flow has been detected, venous pressure is 480 mmHg, and the venous upper pressure limit is 450 mmHg. What are all mandatory machine actions? If none are required, answer: "No mandatory machine action." & Stop the blood pump, stop blood flow, stop anticoagulant flow, and trigger an alarm. \\
During initiation, net fluid removal is enabled, the UFP is rotating backward, and the bypass valve is open. What are all mandatory machine actions? If none are required, answer: "No mandatory machine action." & Stop df flow to the dialyzer, activate bypass mode, and trigger an alarm. \\
During initiation, anticoagulant delivery is running, reverse flow has been detected, and DF temperature is 43 C. What are all mandatory machine actions? If none are required, answer: "No mandatory machine action." & Stop blood flow, stop anticoagulant flow, disconnect the dialyzer from DF flow, activate bypass mode, and trigger an alarm. \\
\midrule
\multicolumn{2}{l}{\textbf{No-Requirement Queries}} \\
During initiation, actual blood flow is 90 mL/min, target blood flow is 200 mL/min, and bypass mode is active. What are all mandatory machine actions? If none are required, answer: "No mandatory machine action." & No mandatory machine action. \\
During initiation, SAD sensor flow is 500 mL/min, detected air volume is 0.4 mL, and the air volume limit is 0.5 mL. What are all mandatory machine actions? If none are required, answer: "No mandatory machine action." & No mandatory machine action. \\
In this machine state, treatment is still ongoing. What are all mandatory machine actions? If none are required, answer: "No mandatory machine action." & No mandatory machine action. \\
\bottomrule
\end{tabular}
\end{table*}

\newpage
\paragraph{Reading the table.} For each requirement flagged as
ambiguity-sensitive in Experiment~6, we report three counts. The
\textbf{Encodings (orig.)} column is the number of distinct
SMT-equivalence classes produced by five independent formalizations of
the \emph{original} requirement: a value of $k$ means the formalizer
read the original requirement in $k$ different ways. The
\textbf{Encodings (final)} column is the same count after the
clarification loop has rewritten the requirement; a value of $1$ means
the formalizer converged on a single reading. The \textbf{$\Delta$}
column is simply \textit{Encodings (final) $-$ Encodings (orig.)} and
reports the size of the convergence effect. All twelve clarified
requirements reach a single encoding class, so the $\Delta$ column is
determined by the original count.

\newpage
\clearpage
\section{Prompt Templates}
\label{app:prompts}

\noindent Placeholders appear as \texttt{\textlangle name\textrangle}.
Model and temperature settings are in Table~\ref{tab:setup}.

\definecolor{promptbg}{HTML}{F7F8FA}
\definecolor{hdrblue}{HTML}{1E3A5F}

\newenvironment{prompt}[1]{%
  \vspace{6pt}
  \noindent{\small\sffamily\bfseries #1}
  \vspace{3pt}
  \begin{center}
  \begin{minipage}{0.92\linewidth}
  \begin{tcolorbox}[
    colback=promptbg, colframe=hdrblue!20, boxrule=0.5pt,
    arc=3pt, left=8pt, right=8pt, top=6pt, bottom=6pt,
    fontupper=\ttfamily\small
  ]
}{%
  \end{tcolorbox}
  \end{minipage}
  \end{center}
  \vspace{2pt}
}

\subsection*{Experiment 1: Autoformalization}

\begin{prompt}{System}
You are a FORMAL METHODS ENGINEER specializing in SMT-LIB encoding.\\
OUTPUT: Pure SMT-LIB v2 only. No prose, no markdown, no code fences.\\[6pt]
REQUIRED STRUCTURE — five sections:\\
\hspace*{1em}; SECTION 1: DECLARATIONS\\
\hspace*{1em}; SECTION 2: DOMAIN CONSTRAINTS\\
\hspace*{1em}; SECTION 3: REQUIREMENTS AS NAMED PREDICATES\\
\hspace*{2em}(define-fun req\_<id> () Bool (=> <pre> <post>))\\
\hspace*{2em}(define-fun req\_<id> () Bool <post>)\\
\hspace*{1em}; SECTION 4: PER-REQUIREMENT CHECKS\\
\hspace*{2em}GUARD-SAT: (echo "GUARD-SAT:<id>") (assert <pre>) (check-sat)\\
\hspace*{2em}VIOLATABLE: (echo "VIOLATABLE:<id>") (assert (not req\_<id>)) (check-sat)\\
\hspace*{1em}; SECTION 5: GLOBAL CONSISTENCY\\
\hspace*{2em}(assert req\_r1) ... (echo "GLOBAL-CONSISTENCY") (check-sat)
\end{prompt}

\begin{prompt}{User}
Produce a complete SMT-LIB v2 file using the structure above.\\[6pt]
REQUIREMENTS:\\
\textlangle requirements\_json\textrangle\\[6pt]
\textlangle schema\_block\textrangle\\[6pt]
Output ONLY SMT-LIB code.
\end{prompt}

\subsection*{Experiment 2: CEGR}

\begin{prompt}{System}
You are answering a hemodialysis machine safety question. State the exact\\
machine actions that should happen. Return exactly one JSON object and nothing\\
else. Every action field must be explicitly true or false.
\end{prompt}

\begin{prompt}{User}
\textlangle variable\_definitions\textrangle\\
\textlangle relevant\_requirements\textrangle\\[4pt]
Scenario: \textlangle scenario\_description\textrangle\\
Question: \textlangle question\textrangle\\[4pt]
Answer using this exact schema: \textlangle response\_schema\textrangle
\end{prompt}

\begin{prompt}{Repair suffix — appended on failure (CEGR condition)}
Your previous response was REJECTED by the formal verifier.\\
Violated requirement(s): \textlangle req\_ids\textrangle\\[4pt]
Counterexample:\\
\textlangle counterexample\_assignments\textrangle\\[4pt]
Conflicts with your previous answer:\\
\textlangle conflict\_lines\textrangle\\[4pt]
Reconsider and produce a corrected JSON response.
\end{prompt}

\noindent{\small The ablation and self-repair conditions use the same base
prompt with the counterexample block removed or replaced by a generic retry
instruction, respectively.}

\subsection*{Exp3 — End-to-End Fault Detection}

\noindent{\small Uses the same system and user prompt as Exp1. The only
difference is that one requirement in \textlangle requirements\_json\textrangle{}
is replaced by its LLM-translated mutant before the prompt is sent.}

\subsection*{Experiment 4: Ambiguity Resolution \& Clarification}

\begin{prompt}{System}
You are a formal methods engineer. Translate natural-language medical-device\\
requirements into SMT-LIB2 assertions. Output only SMT-LIB2 (assert ...)\\
statements. Do not output prose, markdown, comments, or declarations.
\end{prompt}

\begin{prompt}{User}
Variables: \textlangle schema\_variable\_list\textrangle\\[4pt]
Requirement: \textlangle requirement\_text\textrangle\\[4pt]
Output ONLY (assert ...) statements.
\end{prompt}

\noindent{\small The clarification prompt extends the above with solver
evidence (distinct encoding count, semantic entropy, one representative
formula per equivalence class) and instructs the model to rewrite the
requirement minimally until repeated formalization converges. Full template
available in the released code.}

\section{Compute Resources}
\label{app:compute}

All experiments were run on a local CPU workstation using Python and Z3. No
local GPU training or fine-tuning was performed. LLM inference was performed
through hosted API calls to \texttt{claude-sonnet-4-6}; Experiment 2 additionally used
\texttt{claude-haiku-4-5-20251001} for answer-to-assignment translation. Z3 was run locally on CPU.
\newpage
\clearpage

\end{document}